\documentclass[12pt]{article}
\usepackage{graphicx}
\usepackage{amssymb}
\usepackage{amscd}
\usepackage{amsmath}
\usepackage{cite}
\usepackage{slashed}
\begin{document}
\begin{titlepage}

%\begin{center}
%{\hbox to\hsize{
%\hfill \bf hep-ph/??? }}
{\hbox to\hsize{\hfill August 2018 }}

\bigskip \vspace{3\baselineskip}

\begin{center}
{\bf \large 
Charged gravitational instantons: extra CP violation and charge quantisation in the Standard Model}

\bigskip

\bigskip

{\bf Suntharan Arunasalam and Archil Kobakhidze \\ }

\smallskip

{ \small \it
ARC Centre of Excellence for Particle Physics at the Terascale, \\
School of Physics, The University of Sydney, NSW 2006, Australia \\
E-mails: suntharan.arunasalam, archil.kobakhidze@sydney.edu.au }

\bigskip
 
\bigskip

\bigskip

{\large \bf Abstract}

\end{center}
\noindent 
We argue that quantum electrodynamics combined with quantum gravity results in a new source of CP violation, anomalous non-conservation of chiral charge and quantisation of electric charge. Further phenomenological and cosmological implications of this observation are briefly discussed within the standard model of particle physics and cosmology.  
 \end{titlepage}

\section{Introduction}

Gravitational interactions are typically neglected in particle physics processes, because their local manifestations are minuscule for all practical purposes. However, local physical phenomena are also prescribed by global topological properties of the theory. In this paper we argue that non-perturbative quantum gravity effects driven by electrically charged gravitational instantons give rise to a topologically non-trivial vacuum structure. This in turn leads to important phenomenological consequences - violation of CP symmetry and quantisation of electric charge in the standard quantum electrodynamics (QED) augmented by quantum gravity.

Within the Euclidean path integral formalism, quantum gravitational effects result from integrating over metric manifolds $(M, g_{\mu\nu})$ with all possible topologies. The definition of Euclidean path integral for gravity, however, is known to be plagued with difficulties. In particular,  the Euclidean Einstein-Hilbert action is not positive definite, $S_{EH}\lessgtr 0$ \cite{Gibbons:1978ac}.  Nevertheless, for the purpose of computing quantum gravity contribution to particle physics processes described by flat spacetime $S$-matrix , we can restrict ourself to asymptotically Euclidean (AE) or asymptotically locally Euclidean (ALE) manifolds. The AE and ALE vacuum manifolds are known to be Ricci flat, $R=0$, and have non-negative action, $S_{EH}\geqslant 0$, according to the positive action theorem \cite{Schon:1979uj, Witten:1981mf}. Furthermore, while for AE manifolds, $S_{EH}=0$ implies that they are Riemann-flat (no gravity), ALE manifolds with $S_{EH}=0$ nesseccary have (anti)self-dual Riemann curvature tensor. Hence, it is reasonable to think that ALE vacuum manifolds describe a topologically non-trivial vacuum structure of quantum gravity in close analogy to the instanton vacuum structure in a Yang-Mills theory. However, unlike the Yang-Mills instanton background, the background of gravitational (anti)self-dual instantons do not support renormalizable fermion zero modes. This implies that ALE gravitational instantons do not induce e.g. anomalous violation of a global axial charge and are believed have no phenomenological implications in particle physics\footnote{It has been suggested that for global gravitational anomalies the relevant instantons are exotic spheres \cite{Witten:1985xe}. The (non)existence of exotic spheres in 4D, however, has not been proven yet.}. 

The conclusion is dramatically different once one includes into consideration the Standard Model gauge interactions alongside  gravity. Namely, we will argue that electrically charged gravitational instantons support fermion zero modes and hence induce anomalous chiral symmetry breaking in QED. Furthermore, the transition between topologically inequivalent vacua mediated by such instantons give rise to a $\theta$-vacuum and the CP violation in QED. Finally, we will argue that in the background of ALE manifolds that admit spinors, electric charge is necessarily quantised. In addition,  charged gravitational instantons may have important ramifications for cosmology, as it will be briefly discussed at the end of the paper. 

\section{The Eguchi-Hanson instanton}

We start by recalling basic properties of the simplest anti-selfdual gravitational instanton, the Eguchi-Hanson (EH) instanton \cite{Eguchi:1978xp}\footnote{Multi-instanton generation of EH instanton solution is given in Ref. \cite{Gibbons:1979zt}. For a comprehensive review of gravitational instantons,  see  Ref. \cite{Eguchi:1980jx}.}.  
The metric for the EH instanton is:
\begin{equation}
ds^2=\frac{1}{1-\frac{a^4}{r^4}}dr^2+ r^2\left[\sigma_x^2+\sigma_y^2+\left(1-\frac{a^4}{r^4}\right)\sigma_z^2\right]
\end{equation}
where $\sigma_i$ are the differential one-forms:
\begin{align}
\sigma_x=&\frac{1}{r^2}(x \mathop{dt}-t \mathop{dx} +y\mathop{dz}- z \mathop{dy})=\frac{1}{2}(\sin \psi \mathop{d\theta}-\sin \theta \cos\psi \mathop{d\phi})\\
\sigma_y=&\frac{1}{r^2}(y \mathop{dt}-t \mathop{dy} +z\mathop{dx}- x \mathop{dz})=\frac{1}{2}(-\cos \psi \mathop{d\theta}-\sin \theta \sin\psi \mathop{d\phi})\\
\sigma_z=&\frac{1}{r^2}(z \mathop{dt}-t \mathop{dz} +x\mathop{dy}- y \mathop{dx})=\frac{1}{2}(\mathop{d\psi}+\cos\theta \mathop{d\phi})
\end{align}
Defining the curvature 2-form as\footnote{Here and in what follows Greek indices are for curved Eucleadean space, while Latin indices are for tangent (Eucleadian flat) space. Hence, $\gamma^{\mu}={\rm e}^{\mu}_a\gamma^{a}$ are curved space gamma-matrices, $\sigma_{ab}=\frac{i}{4}[\gamma_a,\gamma_b]$  are generators of Eucleadean Lorentz rotations, forming SO(4) symmetry group, and $\omega_{\mu}^{~ab}$ are spin-connection vector fields of the gauged SO(4) symmetry. As usual, spin-connection fields are expressed through tetrad fields ${\rm e}^{a}_{\mu}$ by fulfilling the torsion-free condition. The standard metric formulation then is obtained through the relation: $g_{\mu\nu}=\eta_{ab}{\rm e}^{a}_{\mu}{\rm e}^{b}_{\nu}$.}:
\begin{align*}
R^a_{\ b}=\frac{1}{2}R^a_{\ b\mu\nu}\mathop{dx^\mu}\wedge\mathop{dx^\nu},
\end{align*}
it has been shown in \cite{Eguchi:1978xp} that this metric satisfies the anti-selfduality property:
\begin{equation}
R^a_{\ b}=-\frac{1}{2}\epsilon_{abcd}R^c_{\ d}
\end{equation}
which, in turn, implies that the metric is Ricci flat, $R\equiv R^a_{~a}=0$. \\

The metric is evidently singular at $r=a$ however it can be removed by performing a $\mathbb{Z}_2$ identification of the coordinates. This can be seen as follows.
 Letting $u=r\sqrt{1-a^4/r^4}$, it can be shown that near $r=a$ (or $u=0$), the metric can be rewritten in terms of the Euler angles on $S^3$ as\cite{Eguchi:1978gw}: 
 \begin{align}
ds^2=\frac{1}{4}\mathop{du^2}+\frac{1}{4} u^2 (\mathop{d\psi}+\cos\theta \mathop{d\phi})^2+\frac{a^2}{4}(\mathop{d\theta^2+\sin^2\theta\mathop{d\phi^2}}) 
 \end{align}
Here, it is evident that at fixed $\theta$ and $\phi$, the metric becomes the usual metric of a plane with radial coordinate $u$ and angular coordinate $\psi$. Therefore, to remove the apparent singularity at $u=0$, one must restrict the domain of $\psi$ to $[0,2\pi)$. With this modification, it is clear that the topology of this space near the horizon, $r=a$ is that of $S^2\times \mathbb{R}^2$ where the sphere is parametrised by $(\theta,\phi)$ and the plane by $(u, \psi)$.  As $r\to\infty$, the metric asymptotically approaches that of flat spacetime but with the restriction in the domain of $\psi$, we see that the boundary at infinity is in fact $S^3/\mathbb{Z}_2=\mathbb{R}P^3$.

This metric supports a self-dual  $U(1)$ gauge field (e.g., the electromagnetic field) of the form\cite{Eguchi:1978xp}: 
\begin{align*}
A_r=A_\theta=0\\
A_\phi=\frac{q a^2}{r^2}\cos\theta\\
A_\psi=\frac{q a^2}{r^2}
\end{align*}
where q is the $U(1)$ charge of the instanton. This instanton satisfies the property:
\begin{equation}F^{\mu\nu}=\tilde{F}^{\mu\nu}=\frac{1}{2\sqrt{g}}\epsilon_{\mu\nu\rho\sigma}F_{\rho\sigma}\end{equation}
It should be noted that no such $U(1)$ instanton solution exists in flat space-time and hence the existence of such charged Eguchi-Hanson (CEH) instanton has important phenomenological consequences as seen below. The action of the CEH instanton is given by:
\begin{align}
S_{\rm CEH}=\frac{1}{4e^2}\int d^4x \sqrt{g}F^{\mu\nu}F_{\mu\nu}&=\frac{1}{8e^2}\int d^4x \epsilon_{\mu\nu\rho\sigma}F_{\mu\nu}F_{\rho\sigma} \nonumber \\
&=\frac{4\pi^2 q^2}{e^2}
\label{action}
\end{align}
Note the $4\pi^2$ factor in Eq. (\ref{action}) vs the standard $8\pi^2$ which appears in the action for pure Yang-Mills instantons. This is due identification of antipodal points in the EH space, which becomes a half of the (asymptotic) Euclidean space. More importantly, non-perturbative processes are dominated by small-size CEH instantons due to the growing  fine structure constant $\alpha=e^2/4\pi$ at small scales, in contrast to the dominance of large-size instantons in asymptotically free Yang-Mills theories.

\section{Fermions and their charge quantisation}

In this section we ask the question whether the CEH space actually admits the existence of fermion fields (spin structure). Consider a path at $r=\infty$ from $\psi =0$ to $\psi = 2\pi$ at fixed $\theta$ and $\varphi$. Since the EH space approaches $S^3/\mathbb{Z}_2$ topologically as $r\to \infty$, this path is indeed a loop. Furthermore, this is loop can not be contracted to a point. This can be most easily seen near $r=a$ wherein the space approaches $\mathbb{R}^2\times S^2$ with the sphere parametrised by $\theta$ and $\phi$ and the plane parametrised in plane polar coordinates by $u=r\sqrt{1-a^4/r^4}$ and $\psi$ and the loop goes encloses the origin of the plane. We see that the origin in this plane corresponds to the singularity at $r=a$, thus implying that the loop is not contractible. Considering a fermion with charge $q_e$ moving along such a loop, the additional phase obtained is given by:
\begin{align*}
q_e\oint A_\mu dx^\mu &= q_e\int^\infty_a dr \int^{2\pi}_0 d\psi F_{r\psi}\\
&=4\pi q_e \int_{a}^\infty dr \frac{- q a^2}{r^3}\\
&=-2\pi q_e q
\end{align*}
This accumulated phase must be unobservable for fermion field to be defined consistently \cite{Hawking:1977ab}. Hence, we require $q_eq = n$ for some $n\in \mathbb{Z}.$\footnote{This is reminiscent of the Dirac quantisation condition in the presence of magnetic monopoles \cite{Dirac:1931kp}.}  Since the smallest observed charge carried by down-type quarks is $|q_e|=\frac{1}{3}$ (in units of electron charge), we see that the possible instanton charges are restricted to $q=3n$. We find it quite remarkable that the very existence of fermions in quantum gravity combined with electromagnetism automatically implies quantisation of electric charge.

\section{Anomalous non-conservation of chiral charge and CP violation}

The EH metric by itself forbids the existence of normalisable fermion zero modes even if these are massless. This readily can be seen by squaring the Dirac operator $\slashed{\nabla}\equiv i\gamma^{\mu}(\partial_{\mu}+\frac{1}{4}\omega_{\mu}^{~ab}\sigma_{ab})$:
\begin{equation}
\slashed{\nabla}^2=\nabla_\mu\nabla^\mu-\frac{1}{4}R
\end{equation}
As the EH metric is self-dual, the Ricci scalar, $R$, is zero. Therefore, given a zero mode $\psi$, it also satisfies $\slashed{\nabla}^2 \psi = \nabla_\mu\nabla^\mu \psi =0$. Then, using partial integration, one finds:
\begin{align}
\int d^4x \psi^\dagger \nabla^2\psi=-\int d^4x |\nabla \psi|^2 =0
\label{zero}
\end{align}
where the anti-hermitian property of the covariant derivative was used.  Indeed, Eq. (\ref{zero}) implies that $\nabla_{\mu}\psi=0 \Rightarrow [\nabla_{\mu}, \nabla_{\nu}]\psi =0 \Rightarrow R_{\mu\nu}^{~~ab}\sigma_{ab}\psi=0 \Rightarrow  \psi=0$. Hence, there are no non-trivial normalisable fermion zero modes in the EH space alone.
In other words, $-\nabla_{\mu}\nabla^{\mu}$ is a positive definite operator and has no zero modes.

In contrast, however, if one also introduces the $U(1)$ gauge field, the picture changes significantly. The squared equation operator now becomes: 
\begin{equation}
\slashed{D}^2=\frac{1}{\sqrt{g}}D_\mu\left(\sqrt{g}g^{\mu\nu}D_\nu\right)-\frac{1}{2}i q_e F_{\mu\nu}\sigma^{\mu\nu}
\end{equation}
 where $D_\mu=\nabla_\mu-i q_e A_\mu$. Here, this last operator is indefinite and hence the prior argument fails to hold. In order to illustrate the asymptotic behaviour of this zero mode as $r\to \infty$, consider the following tetrad frame:
\begin{align*}
e^a_{\ \mu}&=\delta^a_\mu +g_-(r) x^\mu x^a + g_+(r)\tilde{x}^\mu \tilde{x}^a
\end{align*}
where $g_\pm (r)=-1+ \left(1-\frac{a^4}{r^4}\right)^{\pm\frac{1}{2}}$ and $\tilde{x}^\mu=(y,-x,t,-z)$.
It is apparent that these tetrads behave asymptotically as $\delta_{\mu a} +O(a^4/r^4) \frac{x_\mu x_a}{r^2}$ and hence the spin connections, which are proportional to the derivative of the tetrads also go as $O(a^4/r^4)$. In contrast, the $U(1)$ field goes as $O(a^2/r^2)$. Hence, we can ignore gravitational effects for $r>> a$ and only consider the $U(1)$ field. In this region, it can be shown that 
$\psi\approx F_{\mu\nu}\gamma^\mu\gamma^\nu\xi_0$ where $\xi_0$ is a constant spinor is a solution of the Dirac equation. This follows from the fact that $\partial_\mu F^{\mu\nu}=0$  and the observation that $\gamma^\mu A_\mu \psi \sim O(a^4/r^4)$ which are considered small in this approximation. This solution goes as $1/r^4$ and hence is normalisable (small-size instantons). 

According to the index theorems \cite{Atiyah:1963zz, Atiyah:1975jf}, the existence of fermion zero modes implies anomalous non-conservation of a chiral charge in gravity-QED mediated nonperturbative processes. The anomalous divergence of the chiral current $J_5^{\mu}=\bar\psi \gamma_5\gamma^{\mu}\psi$ of a fermion $\psi$ carrying a charge $q_e$ reads:
\begin{equation}
\nabla_{\mu}J_5^{\mu}=\frac{q_e^2}{8\pi^2}F_{\mu\nu}\tilde F^{\mu\nu}-\frac{1}{192\pi^2}R^{ab}_{~~\mu\nu}\tilde R^{ab\mu\nu}~.
\label{current}
\end{equation}
By integrating this equation and using Eq. (\ref{action}) we compute change in chiral charge $Q_5\equiv \int d^3x J^0_5$ due to the CEH instanton (the first term on the rhs of \ref{current}), 
\begin{equation}
\Delta Q_5 = 2q_e^2q^2~,
\label{charge}
\end{equation}
while the pure gravitational EH instanton (the second term on the the rhs of \ref{current}) does not contribute. This non-conservation comes in addition to the familiar chiral charge non-conservation due to the QCD instantons, which is believed to be the origin of the $\eta$ meson mass \cite{tHooft:1976snw}. Thus we expect that CEH instantons also contribute to the low energy QCD physics and it would be interesting to study the related phenomenology. The implications of CEH instantons for axion physics is discussed in \cite{Holman:1992ah} and some other interesting phenomenological aspects can be found in \cite{tHooft:1988wxy}.  

Another important phenomenological consequence of CEH instantons is the violation of  CP parity in QED and gravity. Since CEH instantons mediate transitions between topologically distinct vacuum states, the standard cluster decomposition argument implies the following CP violating topological terms are necessarily present in the effective Lagrangian:
\begin{equation}
\frac{1}{\sqrt{g}}{\cal L}_{\slashed{CP}}=\theta_{QED}F_{\mu\nu}\tilde F^{\mu\nu}+ \theta_{grav}R^{ab}_{~~\mu\nu}\tilde R^{ab\mu\nu}
\label{CP}
\end{equation}
The first term is supported by CEH instantons and the second term exists even if $q=0$. Note that in the presence of massless charged fermions $\theta_{QED}$ becomes unphysical and can be removed by chiral transformations, while $\theta_{grav}$ remains, since vacuum-to-vacuum transitions due to the pure EH instantons remain effective due to the absence of fermion zero modes even for massless fermions.       

\section{Outlook: embedding into the Standard Model}

The phenomenological implications of CEH instantons become even richer when one considers the full Standard Model \cite{new}.  The EH instantons charged under the electroweak group $SU(2)\times U(1)$ lead to an anomalous violation of the lepton number in the Standard Model (assuming the absence of right-handed sterile neutrinos) and induce new electroweak CP phases associated with the weak isospin and hypercharge groups. This may have several interesting ramifications which deserve further study. In particular, the electroweak CEH instantons may generate nonperturbative masses for neutrinos, providing instantonic realisation of the gravitational neutrino mass generation mechanism recently suggested in \cite{Dvali:2016uhn}. The simultaneous breaking of CP and lepton number by gravitational instantons could be the source of the observed baryon number (B) asymmetry in the universe. The scenario we keep in mind is that the electroweak CEH instantons (or perhaps equivalent sphalerons), through nontrivial field configurations, induce a lepton number (L) asymmetry at high temperatures. The required departure from thermal equilibrium would be automatically guaranteed, since  that gravitational interactions below the Planck scale cannot sustain in thermal equilibrium. The generated lepton asymmetry is then partly transferred into baryon asymmetry due to the equilibrium B+L number violating processes induced by electroweak sphalerons. 

To conclude, we have argued that charged Eguchi-Hanson gravitational instantons would have a number of important implications for particle physics. Namely, we have identified new CP violating phases and chiral and lepton number violating non-perturbative processes associated with CEH instantons within the Standard Model, without extending its particle content. It also provides a theoretical explanation of the observed quantisation of electric charge of elementary fermions. All this points towards a rather prominent role of quantum gravity in particle physics and cosmology, which has not been fully appreciated previously. 

\paragraph{Note added} After submitting this paper to arXiv, Professors S. Deser and M. Duff have informed the authors about their work \cite{Deser:1980kc} where the possibility of CP violation in QED due to gravitational effects was first suggested.            

\paragraph{Acknowledgements} The authors are indebted to Zurab Berezhiani, Gerard 't Hooft and Arkady Vainshtein for stimulating discussions and Stanley Deser and Michael Duff for their email correspondence. The work was partially supported by the Australian Research Council. AK is also indebted to organisers of the INT Workshop "Neutron-Antineutron Oscillations: Appearance, Disappearance, and Baryogenesis" for the opportunity to present preliminary results of this research.

\end{document}